\documentclass[ aps, prb, amsmath,amssymb,
 reprint,%
]{revtex4-1}

\usepackage{graphicx}
\usepackage{dcolumn}
\usepackage{bm}
\usepackage{epstopdf}

\usepackage{hyperref}
\hypersetup{colorlinks=true,urlcolor=red,citecolor = blue, linkcolor = blue}

\begin{document}

\newcommand{\edit}[1]{\textcolor{green}{\textbf{#1}}}
\newcommand{\ivan}[1]{\textcolor{red}{#1}}
\newcommand{\Rudi}[1]{\textcolor{blue}{#1}}
\newcommand{\figref}[1]{Fig.\,\protect\ref{#1}}
\newcommand{\New}[1]{\textcolor{green}{#1}}

\preprint{AIP/123-QED}

\title{Selective sensitivity in Kerr microscopy}

\author{I.~V.~Soldatov}
\affiliation{Institute for Metallic Materials, Leibniz Institute for Solid State and Materials Research (IFW) Dresden, Helmholtzstrasse 20, D-01069 Dresden, Germany}
\affiliation{Institute of Natural Sciences, Ural Federal University, 620000 Ekaterinburg, Russia}
\author{R.~Sch\"{a}fer}
\affiliation{Institute for Metallic Materials, Leibniz Institute for Solid State and Materials Research (IFW) Dresden, Helmholtzstrasse 20, D-01069 Dresden, Germany}
\affiliation{Institute for Materials Science, TU Dresden, 01062 Dresden, Germany}

\begin{abstract}
A new technique for contrast separation in wide-field magneto-optical Kerr microscopy is introduced. Utilizing the light from eight light emitting diodes, guided to the microscope by glass fibers and being switched synchronously with the camera exposure, domain images with orthogonal in-plane sensitivity can be displayed simultaneously at real-time and images with pure in-plane or polar contrast can be obtained.
The benefit of this new method of contrast separation is demonstrated for permalloy films, a NdFeB sinter magnet, and a cobalt crystal.
Moreover, the new technique is shown to strongly enhance the sensitivity of Kerr microscopy by eliminating parasitic contrast contributions occurring in conventional setups.
A doubling of the in-plane domain contrast and a sensitivity to Kerr rotations as low as 0.6~mdeg is demonstrated

\end{abstract}

\maketitle

\section{Introduction}
\label{introduction}

After the introduction of digital image processing in the 1980ies \cite{SchmidtDiffImg1985}, wide-field Kerr microscopy has become a widely-used and effective tool for magnetic domain imaging. By subtracting an image with domain information from a background image that is free of domains (obtained, e.g., by saturating the specimen), pure domain contrast is obtained in the difference image that can be enhanced by digital means, free of topographic information. Besides general domain imaging on numerous magnetic materials (see Ref.~[\onlinecite{SchaeferBook}] for examples), digitally enhanced Kerr microscopy also provides an easy and direct access to basic physical properties on the microscale of magnetic materials and devices of vital need for spintronic \cite{KehlbergerKerrYIG2015,Gareev2015,MagnusKerrSmCo2014}, spin caloritronic \cite{SoldatovSSE2014} and spin orbitronics \cite{Lavrijsen2014,Hrabec2014,Vanatka2015}. Particulary, if it is combined with other complimentary, integral measurement techniques like magnetometry or transport measurement, the knowledge of the magnetic microstructure gives insight into the origin of the magnetization reversal and related processes. Image processing has furthermore opened various opportunities for Kerr microscopy like quantitative- \cite{Rave1987VectorKerr}, depth selective- \cite{SchaeferDepth1995} and stroboscopic imaging \cite{Flohrer2006}. Reviews on the basics and possibilities of modern Kerr microscopy can be found in Refs. [\onlinecite{SchaeferBook, SchaeferBook2, SchaeferBook3, McCordReview2015}].

\begin{figure}
\center
\includegraphics[width=1\linewidth,clip=true]{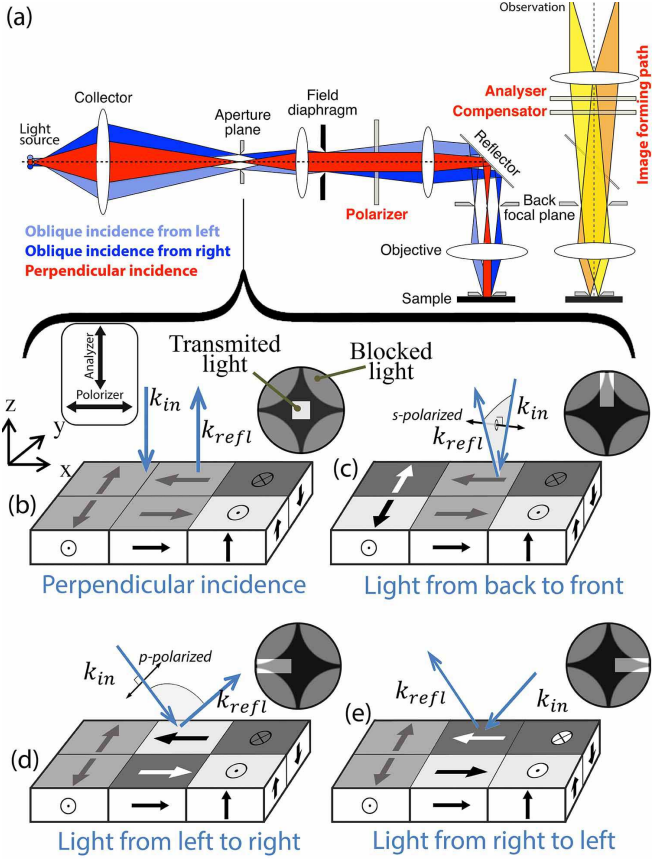}
\caption{
(\textbf{a}) Ray paths for illumination and image formation of a wide-field Kerr microscope with displaceable slit aperture, conventionally used for the adjustment of the sensitivity direction. Beneath the basic geometries of the Kerr contrast are illustrated: polar contrast (\textbf{b}), longitudinal with \textit{s}-polarized light (\textbf{c}), longitudinal in transverse direction with \textit{p}-polarized light with direct (\textbf{d}) and inverted contrast (\textbf{e}). Shown is a perspective view of vertically, horizontally and perpendicularly magnetized domains together with the conoscopic images in which the position of the slit aperture is indicated
}\label{UsualKerr}
\end{figure}

Kerr microscopy utilizes the magneto-optical Kerr effect, i.e. the interaction of plane-polarized light with a non-transparent magnetic media, which in reflection leads to a clock- or counterclockwise rotation of the polarization plane of the incident light depending on the orientation of the magnetization in the media. Often some elliptical polarization in the reflected light is superimposed. The Kerr rotation can be transformed to a domain contrast by placing an analyzer in the reflected light path, and ellipticity can be eliminated with a rotatable compensator (like a quarter-wave plate) in front of the analyzer as indicated in Fig.~\ref{UsualKerr}a. This figure shows separately the illumination and image-formation light paths of a typical wide-field Kerr microscope, which is based on an optical polarization reflection microscope that applies the K{\"o}hler illumination technique to obtain homogeneously illuminated images.
In this scheme the lamp is focused onto the plane of the aperture diaphragm, the light passes through the field diaphragm, is then linearly polarized and deflected downward into the objective lens by a partially reflecting plane glass mirror. After reflection from the sample, the light is collected by the objective and then passes through the half-mirror again before it enters the tube lens, which forms an intermediate image that is further processed towards camera or eyepiece. Since recent years light emitting diode (LED) lamps became sufficiently bright to replace mercury or xenon arc lamps that were traditionally used for Kerr microscopy. LED lamps are highly stable and have a lifetime of several 10.000 hours.

Depending on the mutual orientations of polarization plane, incident light angle and -direction, and magnetization direction in the specimen, the cases of \textit{polar} (magnetization in specimen is out-of-plane) and \textit{longitudinal} (magnetization is in-plane and along the plane of incidence) Kerr effects are distinguishable.
There is also a \textit{transverse} Kerr effect (magnetization in-plane but transversal to the incidence plane) that leads to an amplitude variation rather than a rotation of the reflected light \cite{SchaeferBook,SchaeferBook2}.

From the dielectric law of the Kerr effect, a simple rule can be derived for its symmetry \cite{SchaeferBook2} that immediately leads to the mentioned basic effects: \textit{The Kerr contrast is proportional to the magnetization component along the propagation direction of the reflected light beam}.
This rule is illustrated in Fig.~\ref{UsualKerr}b--e for the most commonly applied magneto-optical effects in Kerr microscopy, the polar and longitudinal Kerr effects.
The polarizer is assumed to be horizontal {along the $x$-axis so that in case of oblique incidence the light is either \textit{s}- or \textit{p}-polarized.
For perpendicular incidence along the $z$-axis (Fig.~\ref{UsualKerr}b), and thus perpendicular reflection, no magnetization components along the reflected light beam do exist for in-plane magnetized domains, which consequently do not show a contrast. Oppositely magnetized out-of-plane domains, on the other hand, have maximum vectorial components, thus showing maximum contrast as expected for the polar Kerr effect.
To get a contrast between in-plane magnetized domains, oblique light incidence is required.
Two cases of the longitudinal Kerr effect are illustrated in the figure: in (c) the light is falling onto the sample surface from the back of the figure with the plane of incidence being vertical along the $y$-axis, leading to a contrast between domains with magnetization components along the plane of incidence.
For pure transverse domains there is no vectorial $M$-component along the reflected light beam $k_\mathrm{refl}$ and thus no contrast.
These domains, however, show up with maximum contrast for light incidence from the left along the $x$-axis (d), whereas the vertical (now transverse) in-plane domains appear without contrast according to the rule.
If the light incidence is inverted along the same axis (image e), the transverse domains again remain without contrast while that of the longitudinal domains is inverted compared to (d).
In all cases of oblique incidence, there are vectorial components of the polar domains along the reflected light beam, so that a polar Kerr effect is always superimposed.
The polar Kerr contrast, however, does not depend on the \textit{direction} of the incident and reflected light, as the magnetization components of polar domains have always the same component along $k_\mathrm{refl}$.
Examples of real domain images for all four cases are shown in Fig.~\ref{SensKerrExmp}.
A Carl Zeiss microscope of the type AxioScope with objective lenses 20x/0.5, 50x/0.8 and 100x/1.3 oil (here the first number is the magnification and the second number the numerical aperture of the lens) was used throughout the paper.

In Fig.~\ref{SensKerrExmp} the plane of incidence and thus Kerr sensitivity was adjusted by placing a slit aperture in the back-focal plane of the microscope as indicated by the circular insets that were already used in Fig.~\ref{UsualKerr}. This "conventional" technique is discussed in more detail in Sect.~\ref{Conventional}, before we will address our novel approach in Sect.~\ref{Advanced}

\begin{figure}
\center
\includegraphics[width=1\linewidth,clip=true]{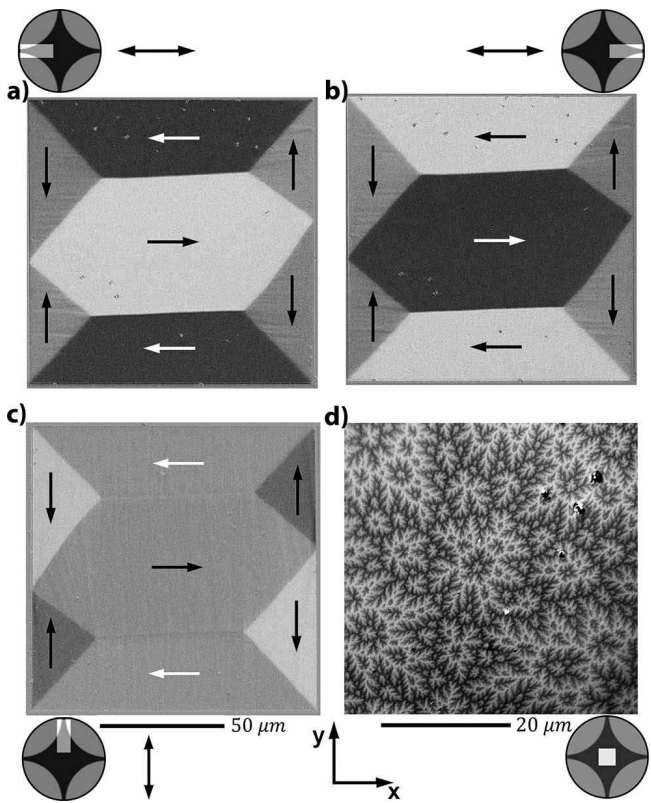}
\caption{
(\textbf{a-c}) Magnetic domains in a Ni$_{81}$Fe$_{19}$ (Permalloy) patterned film element of 240 nm thickness, imaged in longitudinal sensitivities with light incidence from different directions. The contrast with the light coming from the left (\textbf{a}) is opposite to that with the light coming from the right (\textbf{b}). The shown domains reveal a flux-closure pattern, also known as Landau pattern. (\textbf{d}) Out-of-plane magnetic domains at the surface of a cobalt crystal observed on the basal plane and imaged in pure polar sensitivity at perpendicular incidence. Like in Fig.~\ref{UsualKerr}, the direction of incidence is defined by proper positioning the aperture slit as indicated in the conoscopic images. Images (\textbf{a - c}) are difference images in which a saturated background image with the saturation field along the horizontal $x$-axis is subtracted for contrast enhancement, while (\textbf{d}) is a direct image without background subtraction
}\label{SensKerrExmp}
\end{figure}

\section{Conventional Wide-Field Kerr Microscopy}
\label{Conventional}

As discussed in Sect.~\ref{introduction}, the incident light needs to be controlled for the adjustment of a desired Kerr sensitivity.
In a \textit{conventional} wide-field Kerr microscope, the plane of incidence is defined by properly positioning a slit aperture diaphragm in the fully illuminated aperture plane of the microscope, where in case of crossed polarizer, analyzer and compensator a cross-shaped extinction zone can be observed as illustrated in Fig.~\ref{UsualKerr}.
The plane of the aperture is conjugate to the back focal plane of the objective lens, also known as diffraction plane or pupil of the objective, and can be seen in the so-called conoscopical image of the microscope, which is visible by simply looking into the tube after removing the eye piece or by applying a built-in Bertrand lens.
A centered aperture results in effective perpendicular incidence and sole sensitivity to out-of-plane magnetization (polar Kerr effect, see Fig.~\ref{UsualKerr}b). An off-centered aperture diaphragm leads to an obliquely incident bundle of rays as necessary for longitudinal (Fig.~\ref{UsualKerr}c--e) and transverse Kerr sensitivity (not shown).

It has to be noted that approaches are reported in the literature in which the control of light does not require a slit aperture.
In those techniques (see Ref.~\onlinecite{NeudertPy2005} for an example), which often are based on laser illumination for time-resolved Kerr microscopy, the light is guided to the microscope by a glass fiber, the end of which is sitting at the same place as the conventional lamp in Fig.~\ref{UsualKerr}a.
By physically placing the glass fiber end at various positions in the conoscopic plane, different incidence directions can be achieved thus simulating the position of the slit aperture. Such approaches, based on one glass fiber, may also be assigned to the conventional technique.

The conventional Kerr technique, described so far, has a number of shortcomings and problems.
As mentioned, at oblique incidence of light there is always a superposition of sensitivities to in-plane and polar magnetization components.
For the Permalloy film in Fig.~\ref{SensKerrExmp} this is not a problem as this specimen is strictly magnetized in-plane, owing to the fact that for Permalloy the quality factor \cite{SchaeferBook} $Q = K/K_{d}$, which is the relation between anisotropy coefficient $K$ and stray-field energy coefficient $K_{d} = \mu_{0} M_{s}^{2}/2$ (with $M_{s}$ the saturation magnetization), is much smaller than one.
This is different for the cobalt crystal shown in the same figure for which $Q$ is around one.
Here the magnetization strictly follows the perpendicular easy axis in most of the crystal volume. The outermost closure domains, which are finally imaged, however are expected to reveal a stripe-like oscillation of magnetization with polar and in-plane components as was firstly proven by SEMPA (Scanning Electron Microscopy with Polarization Analysis) by the NIST group (see Fig. 2.39 in Ref.~[\onlinecite{SchaeferBook}]).
Because the polar Kerr effect is by a factor of 10 stronger than the longitudinal effect \cite{SchaeferBook2,PolarContrastFootenote}, the Kerr image in Fig.~\ref{SensKerrExmp}d is dominated by the polar magnetization with no indication of the in-plane oscillations.
The separation of in- and out-of-plane magnetization components is thus mandatory to come up with a full understanding of such patterns.
Another example is quantitative Kerr microscopy \cite{Rave1987VectorKerr}, which conventionally only works for strictly in-plane magnetized surfaces due to calibration reasons of the domain contrast.
Contrast separation might extend such analysis to arbitrarily magnetized surfaces.

Another issue is the domain contrast and signal-to-noise ratio that can be improved by a number of approaches like an optimized opening of analyzer and compensator under consideration of the camera sensitivity \cite{SchaeferBook,SchaeferBook3,McCordReview2015}, the mentioned difference-imaging technique \cite{SchmidtDiffImg1985} or normalized differential Kerr microscopy \cite{McCordNormalized1999}.
Nevertheless, for  materials with small Kerr rotations like Permalloy \cite{TanakaKerrAnglePy1963} or thin films of diluted magnetic semiconductors \cite{SoldatovSSE2014,WelpPRL2003} those techniques may be not sufficient and further improvements are desirable to obtain satisfying images.

Finally, as image processing and general domain analysis require some electromagnet around the sample, the applied magnetic field can induce a parasitic Faraday effect \cite{FaradayOrigin1846} in the lenses that are exposed to the field, notably the objective lens.
The Faraday rotation adds to the Kerr contrast and can lead to unwanted domain intensity effects.
By such contributions the quality of domain images may suffer \cite{SoldatovSSE2014}, even lead to a misinterpretation of experimental data \cite{MarkoPaper2015} and bring substantial errors into quantitative Kerr microscopy \cite{RaveVector1993}.

In Sect.~\ref{Advanced} of this article, we report on the development of a new hard- and software realization for contrast separation and enhancement in wide-field magneto-optical Kerr microscopy.
The new technique allows for 3-dimensional vector imaging of the surface magnetization of any specimen, suppresses parasitic Faraday contributions in the lenses for the observation of domains in in-plane configuration, and it leads to a significant enhancement of contrast and signal-to-noise ratio.
Moreover, the developed technique can be used as the basis for a new and advanced technical realization of quantitative Kerr microscopy.

\section{Advanced Design}
\label{Advanced}

The principle of contrast separation is based on the rule for the Kerr contrast mentioned in Sect.~\ref{introduction}. From the schematics in Fig.~\ref{UsualKerr}d,e it is seen that by inverting the direction of oblique light incidence the in-plane contrast is also inverted while the polar contrast remains unaffected.
Thus, by taking the \textit{difference} of the images with inverted light incidence the polar contrast can be eliminated and pure in-plane contrast is left (Fig.~\ref{SensitivityOnSample}, upper panel).
By taking the \textit{sum} of the images, on the other hand, the in-plane contrast is cancelled and pure polar contrast is left (Fig.~\ref{SensitivityOnSample}, lower panel).
Such contrast separation requires that the illumination and contrast conditions for the two directions of incidence are symmetric. If this is not the case (e.g. due to different light intensities), residual unwanted contrast will be left. In a conventional setup using a mechanically adjustable aperture slit for defining the illumination conditions this requirement cannot easily be fulfilled, as the illumination symmetry depends on the precise settings of the slit and a homogenous illumination of the aperture plane.

\begin{figure}
\center
\includegraphics[width=1\linewidth,clip=true]{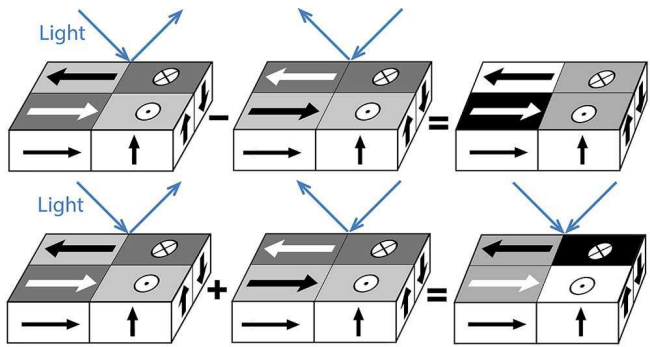}
\caption{
Illustration of contrast separation, based on subtracting or adding two Kerr images that are obtained at opposite directions of light incidence. In the difference image (\textbf{upper panel}) the polar contrast is eliminated, while the in-plane domain contrast is doubled. By adding the images (\textbf{lower panel}), the in-plane contrast is eliminated and pure polar contrast is left. The arrows indicate the magnetization vector at the surface and the areal shading represents the Kerr contrast
}\label{SensitivityOnSample}
\end{figure}

To overcome this shortcoming of conventional Kerr microscopy, a new LED light source was developed that does not require a slit aperture anymore.
It can be connected to any wide-field Kerr microscope that utilizes the K{\"o}hler  illumination scheme (Fig.~\ref{UsualKerr}a).
In this novel setup, the light of eight high-power LEDs is independently guided to the microscope by glass fibers. The fibers have a diameter of 1~mm  each, and the radiant power at the glass fiber output is about 100 mW. The glass fiber ends are physically sitting at the same place as the conventional lamp in Fig.~\ref{UsualKerr}a and they are arranged in a \textit{cross-like} manner.
A similar arrangement was recently published in Ref.~[\onlinecite{HofeVectorKerrFiber2013}], but based on 4 dichromatic LEDs.
The ends of the glass fibers are imaged to the back-focal plane so that they can be seen in the conoscopic image similar to the slit aperture in a conventional setup (compare Figs.~\ref{KerrSetup}c and a).
This corresponds to eight virtual light sources directly in the conoscopic plane.
Such an arrangement allows to set the incidence angle and direction just by turning ON and OFF the appropriate LEDs via computer control without any mechanical action.
Turning ON the LEDs 1\&2 or 3\&4 (Fig.~\ref{KerrSetup}c) corresponds to the opening of the slit for vertical longitudinal (plus polar) sensitivity along the $y$-axis with \textit{p-polarized} light in the conventional arrangement (Fig.~\ref{KerrSetup}a), while 5\&6 or 7\&8 stand for the horizontal longitudinal (plus polar) sensitivity along the $x$-axis with \textit{s-polarized} light.
In the developed light source there is no glass fiber output in the center of the aperture plane, which would be equivalent to the centered slit for the polar contrast as in Fig.~\ref{KerrSetup}b.
The same effect can be achieved, however, if all four LEDs in a row are ON as shown in Fig.~\ref{KerrSetup}d. Then for symmetry reasons (see Fig.~\ref{SensitivityOnSample}, lower panel) the in-plane sensitivity is canceled, providing \textit{pure polar} contrast.
The same is true for all eight LEDs being ON simultaneously.
Note that the number of LEDs, which is imaged to the back focal plane, depends on used objective lens: for low-magnification lenses (like 10x/0.25 or 20x/0.5) all eight fiber ends will be active and visible in the conoscopic image, whereas for high-magnification lenses (like 50x/0.8 or 100x/1.3 oil immersion) with a higher numerical aperture only the inner LEDs (2\&3\&6\&7) may be active. The angle of light incidence covers a range that depends on the numerical aperture of the used objective lens and the respective size of the glass fiber ends in the conoscopic image. For typical lenses the angular ranges of incidence, ÷$\Delta\alpha$, are approximately: (i) lens 20x/0.5: $\Delta\alpha = 2-28^{\circ}$, (ii) lens 50x/0.8: $\Delta\alpha = 11-53^{\circ}$ and (iii) lens 100x/1.3 oil: $\Delta\alpha = 14-59^{\circ}$.

\begin{figure}
\center
\includegraphics[width=1\linewidth,clip=true]{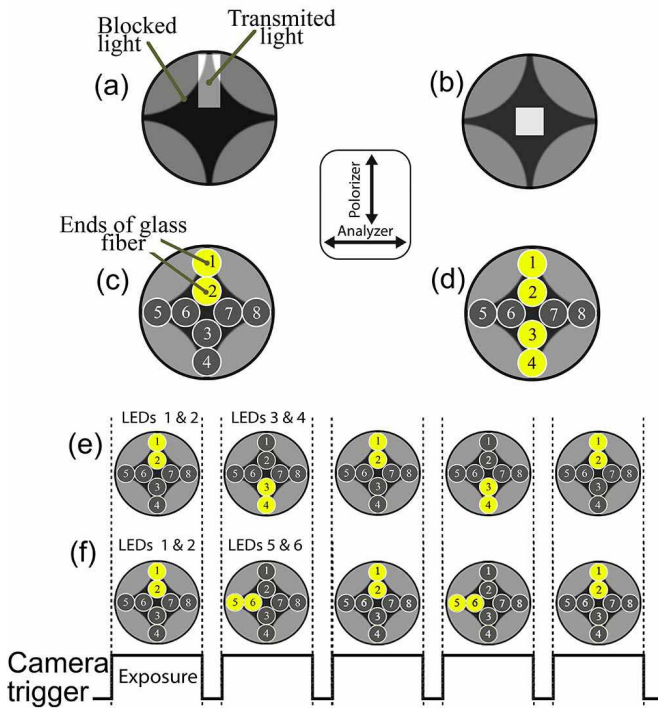}
\caption{
Conoscopic images in a conventional setup with displaced \textbf{(a)} and centered \textbf{(b)} aperture slit for oblique and perpendicular incidence, respectively, and corresponding images for the novel setup \textbf{(c, d)}. The active LEDs are highlighted.
\textbf{(e)} Time diagram of the control sequence for the suppression of polar contrast. The corresponding LEDs are switched ON and OFF synchronously with the camera exposure and the difference of subsequent exposures is taken in-situ. An image with pure in-plane sensitivity is ontaiend then. \textbf{(f)} By alternately activating the LEDs at orthogonal branches of the LED cross leads to dual-component imaging
}\label{KerrSetup}
\end{figure}

With the LED array, the procedure for the suppression of polar sensitivity, based on the principle shown in Fig.~\ref{SensitivityOnSample} (upper panel), is straightforward: the LEDs have to be run in a pulsed mode with a typical switching time in the microseconds range and synchronized with the camera exposure. Two sequential images with opposite incidence angle have to be taken and subtracted from each other as illustrated in Fig.~\ref{KerrSetup}e.
Pure in-plane sensitivity is then achieved.
As long as the frame rate of the camera is beyond approx. 10 frames per second in the triggered mode, this procedure can be applied for real-time imaging of magnetization processes \cite{RealTimeFootenote}.
Also image processing can be applied as for regular, steadily illuminated frames, resulting in high-contrast difference images with pure in-plane sensitivity.
Running all LEDs in a row in a steady mode corresponds to the addition indicated in Fig.~\ref{SensitivityOnSample} (lower panel) and thus leads to pure polar sensitivity as mentioned.

Driving the LED array in the pulsed mode furthermore makes \textit{multi-component imaging} possible: Domain patterns can be imaged under orthogonal sensitivity conditions (e.g. longitudinal vertical and horizontal along the $y$- and $x$-axes, respectively) by alternately activating complementary LEDs on the vertical respective horizontal branches of the cross array as illustrated in Fig.~\ref{KerrSetup}f.
Here polar sensitivity is always superimposed --- pure in-plane sensitivity along the two axes can be achieved by some more sophisticated pulsing schemes, however.
In Ref.~[\onlinecite{HofeVectorKerrFiber2013}] a similar multicomponent Kerr imaging was realized by using LEDs of two different colors along the vertical and horizontal branches of the cross (dichromatic imaging), resulting in two superimposed images with orthogonal sensitivities that can be separated by an image splitter between microscope and camera.
The two complimentary domain images are then displayed in the same frame leading to a bisection and this reduction of the visible sample area.
The two partial images furthermore suffer from different spatial resolutions due to the different wavelengths used to create them.
In our scheme, full frames are displayed for the two complimentary images and they have the same resolution by using monochromatic light of freely selectable color.
The  objection that the pulsed mode suffers from "only nearly simultaneous imaging" \cite{McCordReview2015} turns out to be irrelevant in practice -- static domain images are \textit{averaged} anyway to suppress noise, and the real-time visual observation of magnetization processes doesn't suffer from time lag as long as the frame rate of the camera is faster than approx. 10~fps in the triggered mode.
To obtain pure in-plane sensitivities our pulsed mode can equally well be applied to dichromatic imaging by placing LEDs with one  color on opposite sides of the vertical branch of the LED cross, and diodes with the second color on the horizontal branches, which are then pulsed and triggered according to the described scheme.

As pointed out already, special care has to be payed to the symmetry of the illumination and contrast. The intensities of the light coming from each of the complimentary groups of LEDs have to match in order to avoid residual undesirable sensitivities. For such an equalization, each of the LEDs can be individually dimmed and with a feedback loop, measuring  the integral image intensity at realtime, they can be equalized. Also the contrast \textit{amplitudes} have to match by properly setting analyzer and compensator.

\section{Experimental results}
\label{ExResults}

\subsection{Multicomponent imaging}
\label{PyWall}

An example for dual-component imaging by alternately pulsing LEDs at two orthogonal branches of the LED cross, according to Fig.~\ref{KerrSetup}f, is presented in Fig~\ref{PyWall}.
Here an identical domain pattern of a Permalloy film is shown, imaged at vertical (a) and horizontal (b) sensitivities along the $y$- and $x$-axes, respectively. While image (a) stresses the 180$^{\circ}$ basic domains, a cross-tie wall structure \cite{SchaeferBook} is revealed in (b).
The residual domain contrast in (b) is due to a small misalignement of the easy anisotropy axis that can be verified from the slight tilting of the domain walls away from the vertical axis. If the walls and thus the easy axis would be exactly vertical, there would be no residual domain contrast.
This example demonstrates the necessity of complementary imaging in such "multiaxial" materials to prevent the overlooking of significant information required for a full understanding of the domains.
As this film is largely magnetized in-plane, the inherently superimposed polar sensitivity in this experiment does not play a role.
Exceptions are the cross- and circular Bloch lines \cite{SchaeferBook} sitting at the transitions between black and white wall segments in (b), the cores of which are magnetized perpendicular to the film plane. With a width of some nanometers they are well below resolution, however.

Multicomponent imaging also provides the basis for \textit{quantitative} Kerr microscopy of \textit{complete} magnetization processes.
Like for dichromatic imaging \cite{McCordReview2015}, this can equally well be achieved with our monochromatic technique (to be published elsewhere). Conventionally, quantitative Kerr microscopy can only be applied to static domain images as a mechanical displacement of the aperture slit is required \cite{Rave1987VectorKerr}.

\begin{figure}
\center
\includegraphics[width=1\linewidth,clip=true]{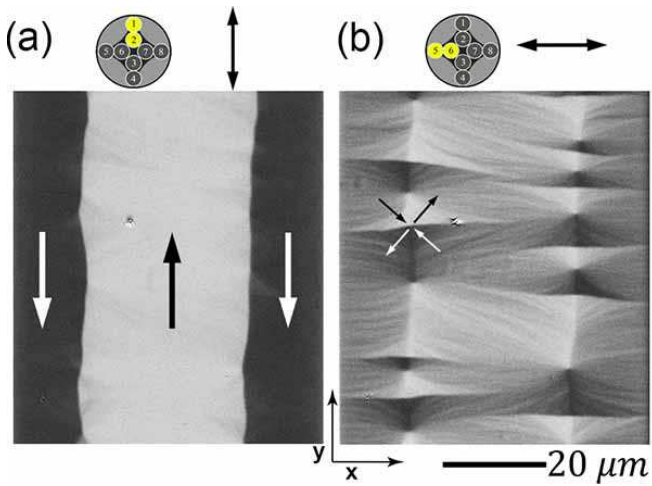}
\caption{
Domain pattern in a 40 nm thick Permalloy film imaged at vertical (\textbf{a}) and horizontal (\textbf{b}) longitudinal sensitivities. The cross-tie walls would be overlooked if only the sensitivity along the walls would have been considered. Shown are difference images for which the background image was taken with an external AC field along the $x$-axis
}\label{PyWall}
\end{figure}

\subsection{Contrast separation}
\label{Contrastseparation}

The possibility of contrast separation with the triggerable LED-based light source is illustrated in Fig.~\ref{NdFeBSeparation}, where it was applied to selective domain imaging on a sintered NdFeB polycrystal.
The magnet was cut so that the texture axis is parallel to the surface along the vertical (y-) direction in the images.
As crystal anisotropy is strong in this material, all domains are strictly magnetized along the easy  \textit{c}-axis that follows the texture axis with some degree of misorientation from grain to grain.
This causes (small) up- and down polar components of magnetization in the individual grains that are superimposed to the predominating in-plane magnetization as schematically shown in the figure.
Under pure polar conditions (Fig.~\ref{NdFeBSeparation}a) only the polar components of magnetization are seen, leading to a complex contrast pattern with different grey levels in the grains that depends on the local misorientation.
If the same domains are imaged under pure in-plane sensitivity (Fig.~\ref{NdFeBSeparation}b), the dominating in-plane domains are showing up in a clear pattern.
As expected, they extend over a number of neighboring grains.
From such pictures the degree of texture can be evaluated which is not easily possible in the polar image.
The same problem would exist if the domains were imaged at oblique incidence in a conventional setup. Then the polar contrast would dominate over the in-plane contrast as the polar Kerr effect is stronger than the longitudinal effect under the given experimental conditions.

\begin{figure}
\center
\includegraphics[width=1\linewidth,clip=true]{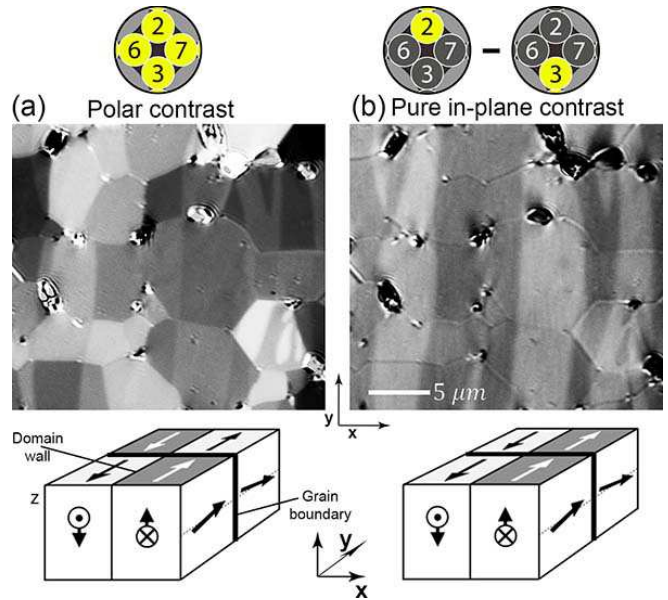}
\caption{
Domains on a sintered NdFeB magnet after thermal demagnetization.  An identical domain state is shown, imaged at pure polar-  \textbf{(a)} and pure in-plane sensitivity \textbf{(b)}. The texture axis is vertical in the domain images. In-plane and perpendicular magnetization components across a grain- and domain boundary are indicated in the sketch. The domain images were obtained without image processing
}\label{NdFeBSeparation}
\end{figure}

A further example of contrast separation is presented in Figure~\ref{CoContrastSepar2}. It is the same branched domain pattern on the hexagonal cobalt bulk crystal that was already discussed in Sect.~\ref{introduction} (compare Fig.~\ref{SensKerrExmp}d). At pure polar sensitivity the domains show up as flower pattern with a high degree of fractal branching (Fig.~\ref{CoContrastSepar2} a). At pure in-plane sensitivity the stripe-like oscillation of magnetization is seen (Fig.~\ref{CoContrastSepar2} b), which has been made visible before with SEMPA \cite{SchaeferBook} as mentioned. Whereas  SEMPA is sensitive to the top most surface magnetization, by Kerr microscopy the information depth is around 20 nm \cite{SchaeferBook2}. So the pure in-plane pattern in Fig.~\ref{CoContrastSepar2}b will be influenced by the magnetization variation within that depth.

In Ref.~[\onlinecite{GutfleischNdFeBKerr2000}, Fig. 3] the separation of perpendicular and in-plane contrasts was already applied to image magnetostatic interaction domains in nanocrystalline NdFeB material. There the contrast separation was still achieved by conventional methodology, i.e. the plane of light incidence was inverted by placing a slit aperture at proper positions in the back-focal plane. A similar contrast separation, applied to magnetic films with tilted anisotropy axis, was reported in Refs.[\onlinecite{McCordReview2015}] and [\onlinecite{UrsDynamics2016}]. In the latter articles the same principle of contrast separation is engaged as in our case, i.e. difference and sum images are formed with LEDs on opposite sides of an LED cross being properly activated. As only two images are processed at a time, contrast separation could only be applied on \textit{static} images. In our case, however, the LEDs are pulsed and triggered together with the camera exposure, so the contrast separation can be achieved on \textit{dynamic} images with a frame rate of at least 10 fms. This is sufficiently fast to observe the contrast-separated images at real-time as fast as the eye can follow. Moreover, as in Refs.[\onlinecite{McCordReview2015}] and [\onlinecite{UrsDynamics2016}] light with different wavelengths was used to obtain the complimentary images, the extraction of the pure in-plane and polar contrasts is questionable due to possible asymmetries in the Kerr rotation constants for the different colors, despite the intensity of light for both incidence directions having been equal. Thus, the claimed estimation of the angle of magnetization from the domain images is dubitable. In our case, LEDs of the same color are used, eliminating the possible problem of a mismatch of the Kerr rotation constants.

\begin{figure}
\center
\includegraphics[width=1\linewidth,clip=true]{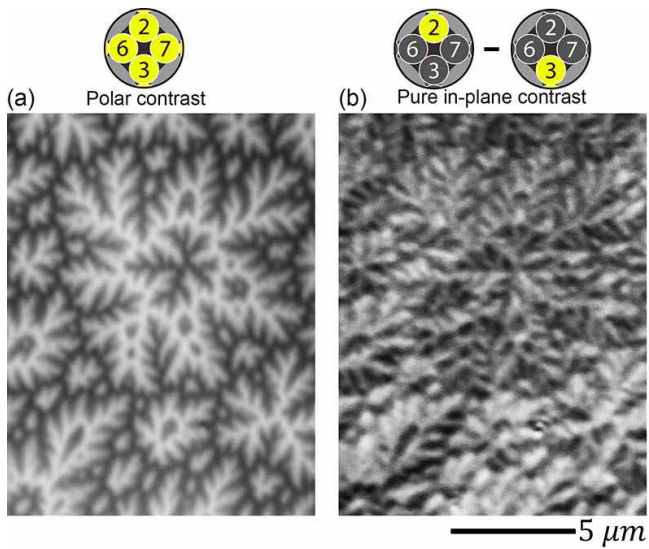}
\caption{
Polar \textbf{(a)} and in-plane \textbf{(b)} magnetization components of an identical domain pattern on a cobalt crystal cut parallel to the basal plane, i.e. with the easy \textit{c}-axis perpendicular to the imaged surface. The domain images were obtained without image processing. For a plot of the magnetization vector field we refer to the quantitative SEMPA images presented in Fig. 2.39 of  Ref.~[\onlinecite{SchaeferBook}]
}\label{CoContrastSepar2}
\end{figure}

\subsection{Faraday correction}
\label{FaradyCorr}

Besides the Kerr effect, there is an additional magneto-optic contribution in a wide field Kerr microscope, the polar Faraday effect that is caused by magnetic field components along the objective lenses as mentioned in Sect.~\ref{introduction}. This effect is significant if perpendicular fields are applied to sample and thus objective. But also inhomogeneous in-plane fields may have polar components along the train of lenses in a typical objective. The Faraday effect manifests itself in a field-dependent intensity that is superimposed to the Kerr intensity of the domains. If magnetization loops are optically measured by plotting the image intensity as a function of magnetic field (MOKE magnetometry), an increasing intensity beyond nominal saturation is a typical indicator of this parasitic effect (see Fig.~\ref{SensitivityKerrCompare} b ahead for an example). Faraday contributions, being mostly linear with respect to the applied field in case of \textit{small} fields, can in principle be eliminated by subtraction of the linear part, for a certain price in accuracy and sensitivity of the experiment, though. The curve in Fig.~\ref{SensitivityKerrCompare} c demonstrates this possibility.
Here the Faraday slope of the curve in (b) was compensated by using a computer routine, leaving a rather noisy MOKE curve. In other cases when the Faraday effect is caused by magnetization-dependent stray-fields of the sample itself \cite{MarkoPaper2015} it is highly non-linear and cannot be easily subtracted from the Kerr-signal. Strong non-linear Faraday effects may also arise in high perpendicular applied field. A hardware separation of Kerr- and Faraday effects is the only option then.

It turns out that running the LED light source in the pure in-plane mode, not only suppresses the polar Kerr contrasts but also the parasitic polar Faraday effect in the objective lens. This possibility was already applied in Ref.~[\onlinecite{MarkoPaper2015}] for domain imaging and MOKE magnetometry on finite-size iron-silicon sheets. Another example is shown in Fig.~\ref{SensitivityKerrCompare} b,f ahead, where magnetization curves are shown that were recorded in a magnetic field swept transverse to the Kerr sensitivity axis.
The dominating Faraday slope in (b) disappears by measuring the loop in the triggered mode of pure in-plane sensitivity (f).

\subsection{Contrast enhancement}
\label{ContrastEn}

Besides the possibilities to compensate the Faraday effect (Sect.~\ref{FaradyCorr}) and separate polar and in-plane magnetization components (Sect.~\ref{Contrastseparation}), the triggered pure in-plane mode also provides a strong enhancement of the sensitivity of the instrument. In Fig.~\ref{SensitivityOnSample} we have seen that the contrast of in-plane domains is inverted for opposite oblique light incidence. If we assign intensities of $+1$ and $-1$ to the bright and dark in-plane domains, respectively, in the conventional, static mode of illumination, then a doubling of the contrast is expected in the difference image of the pure in-plane mode according to $+1-(-1) = 2$ and $-1-(+1) = -2$. In Fig.~\ref{SensitivityOnSample} (upper panel) this is indicated by the gray levels of the sketched domains.

An experimental verification of this contrast enhancement is presented in Fig.~\ref{ContrastEnhancementPy}, where the domains in patterned Permalloy film elements were imaged under regular conditions (LEDs 5\&6 in Fig.~\ref{KerrSetup} are ON) with static oblique illumination (a) and at pure in-plane sensitivity (b) in the pulsed mode (LEDs 5\&6 and 7\&8 being alternatingly triggered). It is clearly seen that the contrast for the latter image is strongly enhanced under the same experimental conditions (exposure time, camera gain, light intensity \textit{etc.}) and without any image post-treatment.

\begin{figure}
\center
\includegraphics[width=1\linewidth,clip=true]{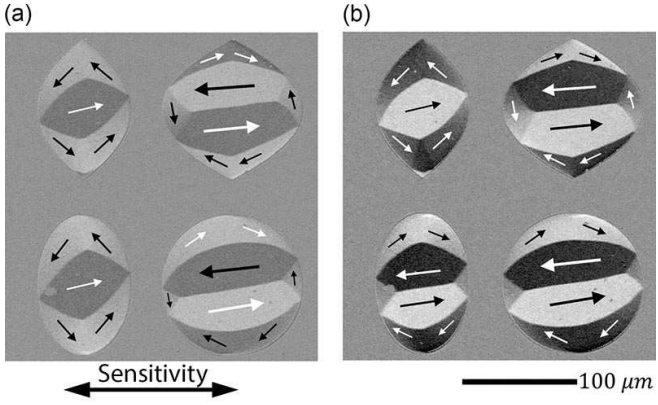}
\caption{
Magnetic domains in Permalloy film elements of 240 nm thickness, taken with horizontal plane of incidence and \textit{p}-polarized light. Image \textbf{(a)} shows a "regular" image in the conventional, static mode. Image \textbf{(b)} was obtained in the pure in-plane (triggered) mode, revealing a doubling of the contrast. Shown are regular difference images (with the saturated state as background image) without any image post-treatment. For both images, 16 frames were averaged to reduce the noise
}\label{ContrastEnhancementPy}
\end{figure}

\subsection{Sensitivity enhancement}
\label{SensitivityEn}

The enhancement of contrast, shown in Sect.~\ref{ContrastEn}, increases the ultimate sensitivity of the measurement system. Figure~\ref{SensitivityKerrCompare} shows optical hysteresis loops, measured on a Permalloy film element in the conventional mode at steady illumination (left) and in the pure in-plane mode (right) without a compensator installed, \text{i.e.} ellipticity components in the reflected Kerr light are not compensated.
In both cases the magnetic field was swept in-plane at an angle $\varphi$ to the direction transversal to the sensitivity axis (see sketch on top of Fig.~\ref{SensitivityKerrCompare}).
At a field angle of $0^{\circ}$, no contrast between the saturated states is thus present.
At a non-zero angle, the projection of the magnetization to the sensitivity axis is non-zero and has different signs for the positive and negative saturation states, so that some contrast between the saturated states will arise.
The peak around zero field, which shows up in the corrected curves, is due to the formation of domains with magnetization along the sensitivity axis.
As can be seen in Fig.~\ref{SensitivityKerrCompare} b, in the conventional mode the Faraday effect in the lenses overwhelms the whole magnetic response of the sample as discussed in Sect.~\ref{FaradyCorr}. After digitally subtracting the Faraday slope (Fig.~\ref{SensitivityKerrCompare} c-e) the remaining Kerr signal occurs to be rather noisy.
In this conventional Kerr mode a distinguishable contrast can be obtained only starting from field angles of approx. $4^{\circ}$ (Fig.~\ref{SensitivityKerrCompare} c,d), while in the pure \textit{in-plane} Kerr mode, due to the absence of the Faraday effect contribution and the enhancement of the signal-to-noise ratio, a contrast can be observed already at $\varphi=1^{\circ}$.
This is also clearly seen from Fig.~\ref{SensitivityKerrCompare}a, where the saturation amplitudes of the loops are plotted as a function of the angle $\varphi$ between external field and direction transversal to the sensitivity axis. The slope in the pure in-plane mode is approximately two times larger than in the conventional mode.

\begin{figure}
\center
\includegraphics[width=1\linewidth,clip=true]{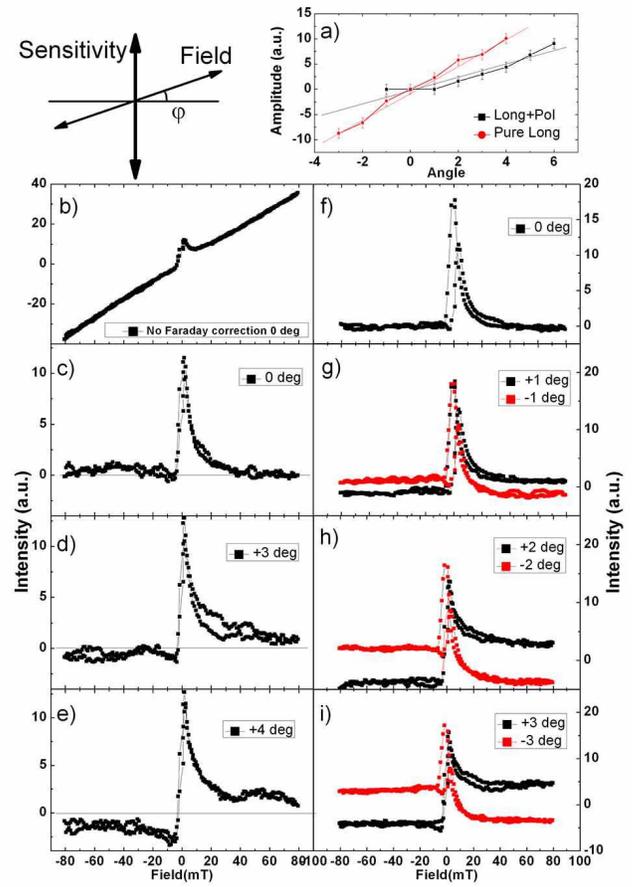}
\caption{
MOKE loops obtained in conventional longitudinal plus polar mode (left panel) and in pure in-plane mode (right panel) on 240 nm thick Permalloy film.
\textbf{(a)} Amplitude of the loops in conventional and in pure in-plane modes.
Loops \textbf{(b, f-i)} are shown as they were measured (before digital subtraction of the Faraday effect contribution) and \textbf{(c-e)} after digital subtraction of the linear part
}\label{SensitivityKerrCompare}
\end{figure}

From these data, a quantitative estimation of the Kerr sensitivity can be obtained as follows:
As the Kerr contrast $C$ is proportional to the Kerr rotation\cite{SchaeferBook}, it would be maximum between the positively and negatively saturated states along the sensitivity axis and proportional to $2\phi_{\text{Py}} $sin$(\varphi)$ \footnote{The factor 2 is due to the geometry of the experiment: the contrast is generated by two oppositely saturated states, which have positive and negative Kerr rotation angles, respectively.} at a any field angle $\varphi$. Here $\phi_{\text{Py}}$ is the longitudinal Kerr rotation angle, which for Permalloy was measured as $\phi_{\text{Py}} \approx 3\cdot10^{-4}$~radians or $30$~mdeg$~=1.8'$ in Refs.~[\onlinecite{TanakaKerrAnglePy1963}] and [\onlinecite{RobinsonPy1963}].
According to Fig.~\ref{SensitivityKerrCompare} we have found that in the conventional mode a distinguishable saturation contrast is obtained for field angles starting at about $\varphi=5^{\circ}$. So the minimal Kerr rotation, that can be clearly detected, is $\phi_{n.m.}= 2\cdot 3\cdot10^{-4} $sin$(4^{\circ})$~radian~$=4 .2\cdot10^{-5}$~radian $=2.4$~mdeg. In the pure in-plane mode it is almost 4 times larger, being as high as $\phi_{p.inp.m.}= 2\cdot 3\cdot10^{-4} $sin$(1^{\circ})$~radian $=1\cdot10^{-5}$~radian $=0.6$~mdeg.
This angle is in the same order as reported for a commercial MOKE magnetometer (NanoMOKE3\textsuperscript{\textregistered} by QuantumDesign \cite{NanoMokefootenote}), with which polarization changes smaller than 0.5 mdeg can be detected in just a few seconds --- in our case some averaging of frames is required that also results in a measuring time of some seconds.


\section{Conclusions}
\label{Conclusions}

In this paper we have presented a novel hard- and software realization for the enhancement of wide-field magneto-optical Kerr microscopy. By pulsing proper LEDs in a cross-shaped arrangement in synchronization with the video camera it is possible (i) to measure and simultaneously display the \textit{x}-and \textit{y}-components of the surface magnetization vector (vectorial Kerr microscopy), which can be used as the basis for a the new technical realization of quantitative Kerr microscopy, and (ii) to separate contrast generated by in-plane and out-of-plane magnetization components of the surface magnetization. Under pure in-plane conditions an enhancement of the in-plane Kerr contrast by a factor of two and the suppression of parasitic Faraday contributions in the objective lens was demonstrated. Due to the  doubling of the contrast, the signal-to-noise ratio is drastically enhanced, which in its turn strongly increases the sensitivity of the method.

\subsection*{Acknowledgement}
\label{Acknowledgement}
Thanks to J. McCord (Kiel) for helpful discussions.

\bibliography{literatura}

\end{document}